\documentclass{article}
\usepackage{amscd,amsmath,amssymb}

\newcommand{\p}[1]{(\ref{#1})}

\newcommand{\cQ}{{\cal Q}}

\newcommand{\cN}{{\cal N}}

\newcommand{\bT}{{\overline T}{}}

\newcommand{\bQ}{{\overline Q}{}}
\newcommand{\bxi}{{\bar\xi}}
\newcommand{\bpsi}{{\bar\psi}}
\newcommand{\beeta}{{\bar\eta}}

\newcommand{\bp}{{\bar p}}
\newcommand{\bz}{{\bar z}}
\newcommand{\bw}{{\bar w}}
\newcommand{\bu}{{\bar u}}
\newcommand{\heta}{\hat{\eta}}
\newcommand{\hxi}{\hat{\xi}}
\newcommand{\hbxi}{\hat{\bar\xi}}

\newcommand{\hbeta}{\hat{\bar\eta}}

\newcommand{\be}{\begin{equation}}
\newcommand{\ee}{\end{equation}}
\newcommand{\bea}{\begin{eqnarray}}
\newcommand{\eea}{\end{eqnarray}}

\newcommand{\ba}{\begin{array}} \newcommand{\ea}{\end{array}}

\newcommand{\im}{{\rm i}}
\newcommand{\diff}{{\rm d}}
\newcommand{\mCP}{\mathbb{CP}}
\newcommand{\mS}{\mathbb{S}}

\newcommand{\nn}{\nonumber}

\def\sfrac#1#2{{\textstyle\frac{#1}{#2}}}

\begin{document}

\thispagestyle{empty}
\setcounter{page}{0}
\setcounter{equation}{0}

\vspace{2cm}
\begin{flushright}
ITP-UH-13/12
\end{flushright}
\vspace{2cm}

\begin{center}
{\LARGE\bf $\cN{=}2$ supersymmetric $\mS^2\to\mCP^3\to\mS^4$ fibration\\[8pt]
viewed as superparticle mechanics }
\end{center}
\vspace{1cm}

\begin{center}
{\Large\bf N.~Kozyrev${}^a$, S.~Krivonos${}^{a}$ and
O.~Lechtenfeld${}^{b}$ }
\end{center}

\begin{center}
${}^a$ {\it
Bogoliubov  Laboratory of Theoretical Physics, JINR,
141980 Dubna, Russia} \vspace{0.2cm}

${}^b$ {\it Institute of Theoretical Physics and
Riemann Center for Geometry and Physics,\\
Leibniz Universit\"at Hannover,
Appelstrasse 2, D-30167 Hannover, Germany}

\end{center}
\vspace{2cm}

\begin{abstract}
We discuss a Hamiltonian reduction procedure that relates the mechanics of an
${\cN}{=}2$ particle on $\mCP^3$ with the motion of such a superparticle
on $\mS^4$ in the presence of an instanton background. The key ingredients of
the bosonic fibration $\mS^2\to\mCP^3\to\mS^4$ are recalled from the viewpoint
of particle mechanics on $\mCP^3$. We describe an ${\cN}{=}2$ supersymmetric
extension which allows for a Hamiltonian reduction. The $\mS^2$ degrees of freedom
are encoded in the supercharges via SU(2) currents. Finally, we present the
Hamiltonian of our system and its superfield Lagrangian.
\end{abstract}

\newpage

\section{Introduction}
The mechanics of a particle moving on a manifold~$E$ is of a special nature if $E$ allows for a fibration,
$F\to E\to B$, i.e.~ if $E$ is fibered over a base manifold~$B$ with fibers~$F$.
If we represent the coordinates of the total manifold~$E$ in terms of base and fiber coordinates,
then the kinetic energy splits into a base part and a fiber part.
If the fibration is nontrivial, e.g., if the total space is not a product of base and fiber,
then particular interaction terms will appear.  These terms are proportional to the particle velocities
and, therefore, describe an interaction with some (possibly non-Abelian) magnetic field.

A prototypical fibration is $\mS^2\to\mCP^3\to\mS^4$, with $\mCP^3$ as total space,
$\mS^4$ as base and $\mS^2$ as fiber.
The problem we address in this letter is the construction of a supersymmetric extension of this
fibration, viewed as a Hamiltonian reduction in $\cN{=}2$ supersymmetric mechanics.
Immediate questions are:
\begin{itemize}
\addtolength{\itemsep}{-6pt}
\item Why do we limit ourselves to $\cN{=}2$ supersymmetry?
\item Does the construction require new features?
\end{itemize}
The first question is easy to answer: We are not aware of any suitable supersymmetric mechanics with $\cN{>}2$
describing particle motion on $\mS^{4k+3}$ and $\mathbb{HP}^k$. The main reason is the absence of a
complex structure on these manifolds, which is crucial for $\cN{=}4$ supersymmetry.\footnote{
The cases of $\mS^2$, $\mS^3$ and $\mS^4 \sim \mathbb{HP}^1$ are rather exceptional,
and the corresponding $\cN{=}4$ supersymmetric mechanics have been constructed in~\cite{s2,s3a,s3b,s4a,s4b}.} 
Keeping in mind a possible extension of our approach to the fibrations $\mS^{4k+3}\to\mCP^{2k+1}\to\mathbb{HP}^k$ 
we limit ourselves to $\cN{=}2$ supersymmetry.

The answer to the second question comes from the paper \cite{Armen1} where the particular fibration
$\mS^2\to\mCP^{3}\to\mathbb{HP}^1$ has been considered in the bosonic case. The crucial ingredient
implicitly used in this paper is the representation of $\mCP^{3}$ as the coset SO(5)$/$U(2) instead of
the commonly used SU(4)$/$U(3) one. This parametrization asked for a special splitting of the standard
complex $\mCP^{3}$ coordinates $\{ z_\alpha, {\bar z}{}^\alpha\,|\; \alpha=1,2,3\}$, transforming under
the fundamental representation of SU(3), into the sets $\{ w_a, {\bar w}{}^a\,|\; a=1,2\}$ and $\{u,{\bar u}\}$,
which form an SU(2) doublet and singlet, respectively, in the coset SO(5)$/$U(2).

Let us recall the construction of \cite{Armen1} in more detail.
Starting from the Lagrangian of a free bosonic particle on $\mCP^3$,
\be\label{bosCP3lag}
{\cal L} _{CP3} = g^{\alpha}{}_{\beta} \dot{z}^{\beta}\dot{\bar{z}}_{\alpha},\qquad
g^{\alpha}{}_{\beta}=\frac{\delta^{\alpha}{}_{\beta}}{1+ z^\gamma \bar z_{\gamma}} -
\frac{z^{\alpha}\bar{z}_{\beta}}{\left(  1+ z^\gamma \bar z_{\gamma} \right)^2},
\ee
and splitting the three complex coordinates $z^{\alpha}$ on $\mCP^3$ into coordinates
on $\mS^4$ and $\mS^2$ via~\footnote{
$\mS^2$ is a complex manifold, thus any of $z^{\alpha}$ coordinates  defines a chart on it.
We choose $z_3$ for this purpose.}
\bea\label{CP3coortrans}
z^1 = \bw^1-\bu\ w_2,\qquad z^2 = -\bw^2-\bu\ w_1,\qquad z^3 = \bu, \nn \\
\bz_1 =w_1-u\ \bw^2,\qquad \bz_2 = -w_2 -u\ \bw^1,\qquad \bz_3=u,
\eea
one may rewrite the Lagrangian \p{bosCP3lag} in the form
\be\label{newbosCP3lag}
{\cal L} _{CP3} = \frac{\dot{w}{}_a \dot{\bw}{}^a}{\left(1+w\cdot\bw \right)^2}+
\frac{\left( \dot{u} -{\cal A} \right)\left( \dot{\bu} -{\cal \bar{A}} \right)}{\left(1+u \bu \right)^2},
\ee
where
\be
{\cal A}=\frac{{w}{}_a \dot{w}{}^a  - u \left( w{}_a \dot{\bw}{}^a - \dot{w}{}_a {\bw}{}^a\right)
+u^2 \bw{}_a \dot{\bw}{}^a}{1+w\cdot\bw}.
\ee
The corresponding Hamiltonian,
\be\label{HPT}
{\cal H} _{CP3}=\left(1+w\cdot\bw\right)^2 P{}^a \bar{P}{}_a + \left( T \bT - U^2\right)
\ee
with
\be\label{newP}
P{}^a = p{}^a + \frac{U\ \bw{}^a  - T\ w{}^a}{1+w\cdot\bw}, \qquad
\bar{P}_a =\bar p{}_a - \frac{U\ w{}_a  - \bT \ \bw{}_a}{1+w\cdot\bw}
\ee
and
\be\label{su2}
T=p{}_u + \bu^2 \bar{p}{}_u, \qquad \bT= \bar{p}{}_u + u^2 {p}{}_u, \qquad U=u{p}{}_u - \bu \bar{p}{}_u,
\ee
displays an interesting feature: the coordinates $\{u,{\bar u}\}$ and corresponding momenta $\{p_u, {\bar p}_u\}$
enter the Hamiltonian only through the SU(2) currents $T,\bT$ and~$U$. The latter satisfy
\be\label{TbTU}
  \{  U ,\ T\}=T,\qquad \{U ,\ \bT \}=-\bT,\qquad \{ T, \bT \}=-2U
\ee
with respect to the standard Poisson brackets
\be\label{PB1}
\{ w{}_a, p{}^b  \}=\delta{}_a^b, \qquad \{ \bw{}^b, \bp{}_a  \}=\delta{}_a^b, \qquad
\{u, p{}_u \}=1, \qquad \{ \bu, \bp{}_u  \}=1.
\ee
This SU(2) is not a symmetry of the system, but the corresponding Casimir operator
\be\label{cassu2}
\mathcal{C}_{su(2)} = T\, \bT - U^2
\ee
Poisson-commutes with the $\mCP^{3}$ Hamiltonian \p{HPT} and, therefore, can be fixed to a constant~$m$.
The currents $T,\ \bT$ and~$U$ do not Poisson-commute with~\p{HPT}, but their brackets are closed, and they fully encode
the $\mS^2$ degrees of freedom in the Hamiltonian. Hence, we may consistently perform the reduction
$\{ T,\;\bT,\;U \}\to 0$ and arrive at a purely $\mS^4$ Hamiltonian.

In what follows, we will construct an $\cN{=}2$ supersymmetric extension of $CP^3$ mechanics,
in which the $\mS^2$ coordinates and momenta enter the supercharges (and hence the Hamiltonian)
only through the SU(2) currents~\p{su2}.
For demonstration we start from the simplest example of $\cN{=}2$ supersymmetric mechanics on $\mS^{2}$.

\setcounter{equation}{0}
\section{$\cN{=}2$ supersymmetric mechanics on $\mS^2$}
\subsection{The standard approach}
In the standard  description of $\cN{=}2$ mechanics on the sphere $\mS^{2}$, which is based on chiral superfields, 
we need, besides bosonic coordinates   $\{u,{\bar u}\}$ and the corresponding momenta $\{p_u, {\bar p}_u\}$,
a pair of fermionic coordinates $\{ \xi, \bxi \}$ subject to the standard brackets
\be\label{pb1}
\left\{ \xi, \bxi\right\} =\im .
\ee
It is straightforward to check that the supercharges
\be\label{Qsu2_st}
Q = \left( 1+ u \bu \right) p_u \bxi, \qquad \bQ =   \left( 1+ u \bu \right) \bp_u \xi,
\ee
form an $\cN{=}2$ super Poincare algebra
\be\label{N2SP}
\left\{ Q ,Q\right\} = \left\{ \bQ ,\bQ \right\} =0, \qquad \left\{ Q, \bQ\right\} =\im H
\ee
with the Hamiltonian
\be\label{Hsu2_st}
H = ( 1+ u \bu )^2 \Bigl( p_u -\im \frac{\bu \xi \bxi}{1+u \bu}\Bigr)
 \Bigl( \bp_u +\im \frac{ u \xi \bxi}{1+u \bu}\Bigr).
\ee
One may check that the supercharges \p{Qsu2_st} and the Hamiltonian \p{Hsu2_st} commute with the SU(2) generators
\be\label{su2_st}
{\tilde T}=p_u+\bu{}^2 \bp{}_u +\im \bu \xi \bxi ,\qquad
\overline{\tilde T}= \bp_u+u^2 p_u-\im u \xi \bxi, \qquad
{\tilde U}= u p_u -\bu \bp_u -\im \xi \bxi
\ee
and that, moreover,
\be
H= \tilde{\mathcal{C}}_{su(2)} ={\tilde T}\, \overline{\tilde T} - {\tilde U}{}^2.
\ee
Clearly the supercharges \p{Qsu2_st} can not be expressed through the SU(2) currents \p{Qsu2_st} alone.
Thus, the standard formulation of $\cN{=}2$ supersymmetric mechanics on $\mS^{2}$ is not suitable for our purposes.

\subsection{Extended version}
The proper and seemingly unique way towards a suitable $\cN{=}2$ supersymmetric mechanics on $\mS^{2}$
extends the number of fermionic variables.
We introduce a second pair of fermionic coordinates $\{\eta, \bar\eta\}$ obeying the canonical brackets
\be\label{eta}
\left\{ \eta, \bar\eta\right\} =\im .
\ee
It is easily checked that the (extended) supercharges
\be\label{Qsu2}
{\cal Q} = T \bar\xi +U \eta -\im \eta \xi \bxi,\qquad
\overline\cQ = \bT \xi -U \bar\eta + \im \bar\eta \xi \bxi
\ee
constructed from the bosonic currents \p{su2}
also generate an $\cN{=}2$ super Poincare algebra \p{N2SP}, but with the purely bosonic Hamiltonian
\be\label{Hsu2}
\sfrac1\im\left\{ \cQ, \overline\cQ \right\} = {\cal H} = \mathcal{C}_{su(2)} = T\, \bT - U^2,
\ee
which is just the bosonic part of~$H$.
This is the version of supersymmetric mechanics we are looking for.

The Hamiltonian \p{Hsu2} resembles the one of $\cN{=}4$ supersymmetric mechanics
on $\mS^{2}$ constructed from a non-linear chiral supermultiplet \cite{s2,Beylin1}. This similarity
is not occasional: one easily finds two additional supercharges
\be\label{Qsu2_extra}
{\cal S} = T \bar\eta -U \xi +\im \xi \eta \bar\eta,\qquad
\overline{\cal S} = \bT \eta +U \bar\xi - \im \bar\xi  \eta \bar\eta
\ee
which commute with the supercharges $\cQ$ and $\overline\cQ$ in \p{Qsu2} and form an $\cN{=}2$ super Poincare algebra
with the same Hamiltonian,
\be\label{Hsu2_ext}
\sfrac1\im\left\{ {\cal S}, \overline{\cal S} \right\} = {\cal H} = \mathcal{C}_{su(2)} = T\, \bT - U^2.
\ee
Thus, we have $\cN{=}4$ supersymmetric mechanics on $\mS^{2}$ exactly the same as in \cite{s2,Beylin1}.
We stress that from the $\cN{=}2$ supersymmetric point of view one needs a general complex $\cN{=}2$ bosonic
superfield (to have four fermions) instead of chiral one more commonly used.

\setcounter{equation}{0}
\section{A novel $\cN{=}2$ supersymmetric mechanics on $\mCP^3$}

To construct an $\cN{=}2$ supersymmetric extension of $\mCP^3$ mechanics with the Hamiltonian \p{HPT}, one
should introduce, besides the fermions $\{\xi,\bar\xi,\eta,\bar\eta\}$ of the previous section,
a doublet of fermionic coordinates $\{ \psi_a ,\bpsi{}^a\}$ with canonical brackets
\be\label{psi}
\left\{ \psi_a, \bpsi{}^b \right\} = \im \delta_a^b, \qquad \left( \psi_a \right)^\dagger = \bpsi{}^a .
\ee
These fermionic coordinates accompany the bosonic coordinates $\{w_a, {\bar w}{}^a\}$
to render our system $\cN{=}2$ supersymmetric.

We are ready to present the supercharges,
\bea\label{Qsp3}
Q = \left( 1+ w\cdot \bw\right) P^a \psi_a -\im\, \psi_a \bw{} ^a\, \psi_b \bpsi{}^b +
\alpha\left( T \bxi+ U\eta-\im\, \eta \,\xi \bxi\right)+
 \im  \psi_a w^a \eta \xi -\sfrac{\im}{\alpha}  \psi_a \psi^a \xi -\im \psi_a \bw{}^a \xi\bxi,&&
\nn \\[4pt]
\bQ = \left( 1+ w\cdot \bw\right) {\overline P}_a \bpsi{}^a + \im\, w_a \bpsi^a \, \psi_b \bpsi{}^b +
\alpha\left( {\overline T} \xi- U{\bar\eta}+\im\, \bar\eta \,\xi \bxi\right)-
 \im  \bw_a \bpsi^a  \bar\eta \bxi -\sfrac{\im}{\alpha}  \bpsi_a \bpsi^a \bxi +\im w_a\bpsi^a \xi\bxi,&&
\eea
where the momenta $\{P^a, {\overline P}_a \}$ were defined in \p{newP}, and $\alpha$ is a real parameter.
Its sign is irrelevant, due to the involution
$\{\eta,\bar\eta,\xi,\bxi\}\to \{-\eta,-\bar\eta,-\xi,-\bxi\}$ which maps $Q(\alpha)$ to $Q(-\alpha)$.

The supercharges \p{Qsp3} commute as needed,
\be
\left\{Q, Q \right\} =\left\{ \bQ, \bQ \right\}=0, \qquad \left\{ Q, \bQ \right\} = \im H ,
\ee
with the Hamiltonian
\be\label{HamCP}
H=  \left( 1+ w\cdot \bw\right)^2 \left(P^a-\im A^a\right)\left( {\overline P}_a+\im {\overline A}_a\right) +
\alpha^2 \left( T+\sfrac{\im}{\alpha} B\right)\left(\bT-\sfrac{\im}{\alpha}{\overline B}\right) -
\alpha^2 \left(U+\sfrac{\im}{2\alpha^2} B_u \right)^2+H_{4f}.
\ee
Here,
\bea\label{connections}
A^a&=&\frac{1}{1+w\cdot \bw}\bigl( \sfrac{2}{\alpha}\psi^a \xi+w^a \xi\eta +\bw^a( \xi\bxi+\psi_b\bpsi^b)-
2 \bw^b\psi_b \bpsi^a\bigr), \nn\\
{\overline A}_a&=&\frac{1}{1+w\cdot \bw}\bigl( \sfrac{2}{\alpha}\bpsi_a \bxi+\bw_a \bxi\bar\eta +
w_a ( \xi\bxi+\psi_b\bpsi^b)+ 2 w_b\bpsi^b \psi_a\bigr), \nn\\[4pt]
B&=& \bw^a \bpsi_a \eta,\qquad {\overline B}=w_a \psi^a \bar\eta, \qquad
B_u= 2 \psi_a \bpsi^a -\alpha w_a \psi^a \xi -\alpha \bw^a\bpsi_a \bxi,
\eea
and the four-fermion term takes the form
\bea\label{4f}
H_{4f}&=&\left(\alpha w_a\psi^a\xi+\alpha\bw^a\bpsi_a\bar\xi+2w_a\bw^a\psi_b\bpsi^b-
2\bw^a\psi_a w_b\bpsi^b\right)\eta\bar\eta+ \nn\\
&& \left(\sfrac{3}{2} \bw^a\psi_a w_b\bpsi^b-\sfrac{3}{2} w_a\bw^a \psi_b\bpsi^b-2 \psi_a \bpsi^a\right) \xi\bxi +\nn\\
&&\left(-\xi\bar\eta-\sfrac{3}{2\alpha} w_b\bpsi^b\xi +\sfrac{1}{\alpha}\bw^b\bpsi_b \bar\eta\right)\psi_a\psi^a+
\left( \bxi\eta +\sfrac{3}{2\alpha} \bw^b\psi_b\bxi+\sfrac{1}{\alpha}w_b\psi^b\eta\right) \bpsi^a\bpsi_a+ \nn\\
&& + \sfrac{1}{\alpha^2}(1-2\alpha^2+\alpha^2 w\cdot \bw)\, \psi_a\bpsi^a\, \psi_b\bpsi^b .
\eea
The bosonic part of the Hamiltonian $H$ \p{HamCP} reads
\be\label{Hcp3_bos}
H_{bos} = \left( 1+ w\cdot \bw\right)^2 P^a {\overline P}_a +\alpha^2 \left( T \bT -U^2 \right).
\ee
For $\alpha{=}1$ this Hamiltonian coincides with the Casimir operator of the SU(4) group and, therefore, with
the Hamiltonian of bosonic $\mCP^3$ mechanics. Another value, $\alpha{=}\sqrt{2}$, corresponds to the
Casimir operator of~SO(5).

Clearly, the reduction
\be\label{red}
\left\{ T, \bT, U , \eta, \bar\eta, \xi,\bxi\right\}\ \rightarrow\ 0
\ee
proceeds nicely and yields the reduced supercharges
\be\label{Qso4}
Q_{red} = \left( 1+ w\cdot \bw\right) p_w^a \psi_a -\im\, \psi_a \bw{} ^a\, \psi_b \bpsi{}^b ,\qquad
\bQ_{red} = \left( 1+ w\cdot \bw\right) {\overline p}_w{}_a \bpsi{}^a + \im\, w_a \bpsi^a \, \psi_b \bpsi{}^b .
\ee
They properly commute to the reduced Hamiltonian
\be
\sfrac1\im\left\{ Q_{red}, \bQ_{red} \right\} = H_{red} =
\left( 1+ w\cdot \bw\right)^2 \left(p_w^a-\im {\hat A}{}^a\right)\left( {\overline p}_w{}_a +\im {\overline{\hat A}}_a\right)+
\left(-2+w\cdot \bw\right)  \, \psi_a\bpsi^a\, \psi_b\bpsi^b
\ee
with
\be
{\hat A}{}^a=\frac{1}{1+w\cdot \bw}\left( \bw^a \psi_b\bpsi^b-2 \bw^b\psi_b\bpsi^a\right),\qquad
{\overline{\hat A}}_a=\frac{1}{1+w\cdot \bw}\left( w_a \psi_b\bpsi^b+2 w_b\bpsi^b\psi_a\right).
\ee
We have produced a $\cN{=}2$ supersymmetric extension of mechanics on $\mS^4$.

\section{Superfield description}

\subsection{Component Lagrangian}

Before constructing a superfield action for the system presented, it is instructive to find the component Lagrangian
from our Hamiltonian \p{HamCP} at $\alpha{=}1$. This can be done by performing a Legendre transformation
over the bosonic variables and by adding fermionic kinetic terms which produce the Dirac brackets \p{pb1}, \p{eta} and \p{psi}.
The resulting Lagrangian reads
\bea\label{LagrCP}
L = \frac{\dot w{}_a \dot{\bw}{}^a}{(1+w\cdot \bw)^2} +
\frac{\im}{2} \bigl( \dot\psi{}_a \bpsi{}^a -  \psi{}_a \dot\bpsi{}^a  \bigr)+
\frac{\im}{2} \bigl( \dot\xi \bxi -  \xi \dot\bxi  \bigr)+
\frac{\im}{2} \bigl( \dot\eta \beeta -  \eta \dot\beeta  \bigr) +
\nn \\
+ \im A{}^a \dot w{}_a -\im \bar{A}{}_a \dot \bw{}^a + \frac{(\dot u + \Lambda)(\dot\bu + \bar\Lambda)}{ (1+u\bu)^2}-
B\bar B -\frac{B_{u}^2}{4} - H_{4f},
\eea
where
\bea
&&\Lambda = \frac{(w{}^a-u\bw{}^a)\,\dot{w}{}_a + (u w{}_a - u^2 \bw{}_a)\,\dot{\bw}^a}{1+w\cdot\bw} -
\im \left( B u^2 - \bar B - u B_{u}   \right), \nn \\
&&\bar\Lambda= \frac{(\bu^2 w{}^a+\bu\bw{}^a)\,\dot{w}{}_a - (\bu w{}_a + \bw{}_a)\,\dot{\bw}^a}{1+w\cdot\bw} -
\im \left( B  - \bar B \bu^2 + \bu B_{u}   \right).
\eea
Since the key ingredient for the superfield description of the Lagrangian \p{LagrCP} is
a superfield Lagrangian for the extended version of
$\cN{=}2$ mechanics on $\mS^2$ discussed in Subsection 2.2, we begin with this case.

\subsection{Superfield action for $\mS^2$}
As we already stressed, chiral superfields are not sufficient for our case.
Thus we start from two general $\cN{=}2$ superfields $\{u,\bu\}$ with an ansatz for the action,
\bea\label{S2superfac}
S_{S2}=-\int\!\diff{t}\diff\theta\diff\bar\theta \left( F_1 Du \bar D \bu + F_2 D\bu \bar D u   \right).
\eea
The superfields  $\{u,\bu\}$ are unconstrained and, therefore, contain the following components,
\be\label{S2comp}
u = u|, \quad \bu = \bu|, \quad
\hxi=i\bar D \bar u|, \quad  \hbxi=iDu|, \quad \heta=i D \bu |, \quad \hbeta=i\bar D u |, \quad
A=\sfrac12 \left[D, \bar D \right]u|, \quad \bar A=\sfrac12 \left[ D, \bar D \right] \bu|,
\ee
where, as usual, $|$ denotes the $\theta,\bar\theta\to 0$ limit.

Our task is to fix the arbitrary functions $F_1$ and $F_2$ entering \p{S2superfac} in such a way as to get
the proper component Lagrangian~\footnote{
This Lagrangian follows from \p{LagrCP} in the limit $\{w_a,\psi_a\}\to 0$.}
\be\label{L3}
{\cal L} _{S2}= \frac{\dot u \dot \bu}{(1+u\bu)^2} + \frac{\im}{2}\bigl( \dot\xi \bxi - \xi  \dot\bxi   \bigr) +
\frac{\im}{2}\bigl( \dot\eta \bar\eta - \eta  \dot{\bar\eta}   \bigr)
\ee
after integration over $\theta$ and $\bar\theta$ in \p{S2superfac} and elimination of the auxiliary components
$A$ and ${\bar A}$~\p{S2comp}.
Finally, one has to determine the relation between $\{\hxi,\hbxi,\heta,\hbeta\}$ and $\{\xi,\bxi,\eta,\bar\eta\}$.
Comparing the transformations properties of our superfields $\{u, \bu\}$
to those which follow from our explicit construction of supercharges in \p{Qsu2}, one finds that
\be\label{S2ferm}
\hxi=\sqrt{2}( \xi + \bu \beeta), \qquad \hbxi =\sqrt{2}( \bxi +  u\eta),  \qquad
\heta=-\sqrt{2}(\bu\eta - \bu^2 \bxi), \qquad \hbeta=\sqrt{2} (u \beeta - u^2 \xi).
\ee

After integration over $\theta$ and $\bar\theta$ in \p{S2superfac} and eliminating the auxiliary fields
by their equations of motion, we arrive at the Lagrangian
\bea\label{lag2}
{\cal L} _2 = \Bigl[  F_1 + F_2 -\frac{(F_1 - F_2)^2}{F_1 + F_2}\Bigr] \dot u \dot\bu +
\im F_1 \bigl(  \dot \hxi \hbxi - \hxi \dot \hbxi \bigr) + \im F_2 \bigl(  \dot \heta \hbeta -  \heta \dot \hbeta \bigr)-
\im (\dot u \bu - u \dot \bu ) \bigl( F^{\prime}_1 \hxi \hbxi + F^{\prime}_2 \heta \hbeta \bigr)+
\nn \\
+\bigl( F^{\prime }_1  + u \bu F^{\prime\prime}_1 + F^{\prime }_2  + u \bu F^{\prime\prime}_2 \bigr) \hxi \hbxi \heta\hbeta +
\im\Bigl[ F^{\prime}_1 - F^{\prime}_2 - \frac{F_1 - F_2}{F_1 + F_2} (F^{\prime}_1 + F^{\prime}_2 ) \Bigr]
\bigl( u\dot u \heta\hxi - \bu \dot \bu \hbxi \hbeta \bigr).
\eea
After passing to the proper fermions via \p{S2ferm}, one may easily check that the Lagrangian \p{lag2}
coincides with  \p{L3}, if we will fix our functions $F_1$ and $F_2$ as
\be\label{func12}
4\,F_1 = \frac{1}{1+u\bu}, \qquad 4\,F_2 = \frac{1}{u\bu\,(1+u\bu)}.
\ee
This proves that the superfield action \p{S2superfac} indeed yields the component Lagrangian \p{L3}
upon the choice~\p{func12} for $F_1$ and $F_2$.

How unique is this result?
One may wonder about adding further terms to the superfield action \p{S2superfac}, such as
\be\label{uselterm}
F_3\,u^2 D\bu \bar D \bu, \qquad F_4\,\bu^2 D u \bar D u, \qquad \im F_5\,(\dot u \bu - \dot \bu u).
\ee
A detailed analysis shows that these terms do not play any role: to reproduce~\p{L3},
the functions $F_3$ and $F_4$ have to vanish while the $F_5$ term can be absorbed into \p{lag2}
by a suitable modification of $F_1$ and~$F_2$.
Hence, the superfield action \p{S2superfac} is completely fixed.

\subsection{Complete superfield action}
Having at hands the superfield Lagrangian for our $\cN{=}2$ mechanics on $\mS^2$,
it is straightforward to construct the full superfield Lagrangian for the complete system on $\mCP^3$,
\be\label{LagrCPspf}
{\cal L}= -\frac{1}{4}\biggl[\frac{Dw{}_a\bar D\bar w{}^a}{\left(1+w\cdot\bar w\right)^2}+\frac{(Du+M^a Dw{}_a)(\bar D\bar u+
\bar M_a \bar D \bar w{}^a)}{1+u\bu}+\frac{(D\bu+N^a Dw{}_a)(\bar D u+\bar N_a \bar D \bar w{}^a)}{u\bu(1+u\bu)}\biggr]
\ee
with
\be\label{MN}
M{}^a = \frac{w{}^a  - u \bar w{}^a}{1+w\cdot \bar w}, \qquad N{}^a = \frac{\bu^2 w{}^a  + \bu \bar w{}^a}{1+w\cdot \bar w}.
\ee
Coincidence with the full component Lagrangian \p{LagrCP} is achieved upon the following identification
of the physical components in our superfields,
\be\label{comps}
\hxi=\im(Du+M^a Dw{}_a )\vert , \qquad  \heta = \im(D\bu+N^a Dw{}_a)\vert ,\qquad
\psi{}_a =\frac{\im Dw{}_a}{\sqrt{2}(1+w\cdot\bw)} \vert ,
\ee
and subsequent redefinitions \p{S2ferm}.

\setcounter{equation}{0}
\section{Conclusion}
In this Letter we have proposed a Hamiltonian reduction procedure that relates the mechanics of an
$\cN{=}2$ particle on $\mCP^3$ with the motion of such a superparticle
on $\mS^4$ in the presence of an instanton background.
The key ingredient for the existence of the $\cN{=}2$ extension of the bosonic reduction
was a novel action for the $\cN{=}2$ supersymmetric particle moving on $\mS^2$.
It turned out that the latter is nothing but $\cN{=}4$ supersymmetric mechanics with a non-linear
chiral supermultiplet, written in terms of $\cN{=}2$ superfields. The supercharges contain the bosonic
variables and their momenta only via SU(2) currents. Therefore, the Hamiltonian reduction procedure
paralleled the purely bosonic one. The full system, for which we have constructed the Hamiltonian
and supercharges, is a new variant of $\cN{=}2$ mechanics for a superparticle on $\mCP^3$.
It contains twice as many fermions as the standard version based on $\cN{=}2$ chiral superfields.
We also provided a superspace description for our system.

Our construction may be extended to reductions over $\mS^3$ as well as
to  $\mS^{4k+3}\to\mCP^{2k+1}\to\mathbb{HP}^k$ fibrations.
We will consider these extensions elsewhere.

\section*{Acknowledgements}
We are indebted to Armen Nersessian  for valuable discussions.
This work was partly supported by Volkswagen Foundation grant I/84 496, by RFBR grants  11-02-01335-a,
 11-02-90445-Ukr, 12-02-00517-a.

\end{document}